# Experimental demonstration of three-color entanglement

*Three bright light beams of different colors are entangled and the robustness of this resource under a lossy channel is investigated.*

A. S. Coelho[1], F. A. S. Barbosa[1], K. N. Cassemiro[2], A. S. Villar[2,3], M. Martinelli[1], P. Nussenzveig[1*]

[1]Instituto de Física, Universidade de São Paulo, P. O. Box 66318, São Paulo, SP 05314-970, Brazil. [2] Max Planck Institute for the Science of Light, Günther Scharowskystr. 1 / Bau 24, 91058 Erlangen, Germany. [3]Universität Erlangen-Nürnberg, Staudtstr. 7/B2, 91058 Erlangen, Germany.

To whom correspondence should be addressed. E-mail: nussen@if.usp.br

**Entanglement is the essential quantum resource for a potential speed-up of information processing, as well as for sophisticated quantum communication. Quantum information networks will be required to convey information from one place to another, by using entangled light beams. Many physical systems are under consideration as building blocks, with different merits and faults, so that hybrid systems are likely to be developed. Here we present an important tool for connecting systems that share no common resonance frequencies: we demonstrate the first direct generation of entanglement among more than two bright beams of light, all of different wavelengths (532.251 nm, 1062.102 nm, and 1066.915 nm). We also observe, for the first time, disentanglement for finite channel losses, the continuous variable counterpart to entanglement sudden death.**

Ever since Schrödinger presciently defined entanglement as "the characteristic trait of quantum mechanics" (*1*), much study has been devoted to its understanding. Possible applications to information tasks such as storage, computation, and communications, have generated a large amount of research (*2*). Quantum computing is expected to be more efficient than its classical counterpart, and there are quantum algorithms (*3*) that greatly surpass the best classical ones known. Quantum communications can, in principle, provide absolute security (*4*). Nevertheless, many technological challenges remain, since the quantum resources are fragile and undergo decoherence from the inevitable interactions with the surrounding environment (*5*). Furthermore, many fundamental issues remain to be understood, regarding the nature and dynamics of entanglement (*6*).

A variety of physical systems are currently under investigation to perform the envisioned information tasks (*7–12*). Since each one has its advantages and disadvantages, it is likely that several of them will be combined for reliable quantum information tasks. Light, in view of its high speed and weak interaction with the environment, is the natural candidate to convey quantum information among them by using quantum teleportation (*13,14*). In order to connect such different physical systems at the nodes of a quantum network (*15*), different frequencies of light will be necessary. For two such beams, entanglement has already been demonstrated, ranging from small frequency differences (*16,17*) up to one frequency being twice the other (*18*). It is important to go beyond just two beams for multimode networks. Our demonstration employs three field modes with different wavelengths, or colors.

Our system is the optical parametric oscillator (OPO), one of the most studied systems in quantum optics. It consists of a non-linear optical crystal inside a cavity, so that the fields are fed-back into the system. Light incident on the crystal undergoes parametric down-conversion, a process in which an incident photon is converted into a pair of generated photons, such that there is energy conservation:

$$\omega_0 = \omega_1 + \omega_2 , \qquad (1)$$

where $\omega_i$ ($i=0,1,2$) are the angular frequencies of the incident light and of the generated photons, respectively. Momentum is also conserved, corresponding to the phase matching condition. Owing to the cavity feedback, down-converted photons can be emitted in occupied field modes, a process known as stimulated emission, with increasing probability as the number of photons in the mode grows. Gain is thus obtained. By increasing the pump laser power, the gain overcomes the losses and the system oscillates. Above this oscillation threshold, tripartite entanglement is predicted ([19]).

Tripartite entanglement follows from eq. (1). On the one hand, down-converted photons are produced in pairs, yielding strong intensity correlations among the so-called twin beams. In order to produce twin photons, a pump photon must be annihilated, thus anti-correlations are expected between the reflected pump intensity and the sum of twin beams' intensities. On the other hand, the frequency constraint of eq. (1) translates into a constraint for the phase variations (or fluctuations) of the three fields. The twins' phase fluctuations should be anti-correlated and their sum should be correlated to the pump's phase fluctuations. A criterion for analyzing tripartite entanglement, by van Loock and Furusawa ([20]), is written directly in terms of these correlations:

$$V_0 = \Delta^2\left(\frac{p_1 - p_2}{\sqrt{2}}\right) + \Delta^2\left(\frac{q_1 + q_2}{\sqrt{2}} - \alpha_0 q_0\right) \geq 2$$

$$V_1 = \Delta^2\left(\frac{p_0 + p_1}{\sqrt{2}}\right) + \Delta^2\left(\frac{q_0 - q_1}{\sqrt{2}} - \alpha_2 q_2\right) \geq 2 \quad . \quad (2)$$

$$V_2 = \Delta^2\left(\frac{p_0 + p_2}{\sqrt{2}}\right) + \Delta^2\left(\frac{q_0 - q_2}{\sqrt{2}} - \alpha_1 q_1\right) \geq 2$$

It suffices to violate two of these inequalities to demonstrate tripartite entanglement. The $p_i$ ($i=0,1,2$) represent the amplitude quadratures of the fields, the $q_i$ ($i=0,1,2$) represent their phase quadratures and the $\alpha_i$ are free parameters, chosen to minimize the inequalities. Phase and amplitude quadratures of each field do not commute, leading to a minimum uncertainty product, which characterizes the Standard Quantum Limit (SQL). Variances in Eq. (2) are normalized to the SQL.

The experiment is designed to measure general quadratures of the three fields. The setup is sketched in Fig. 1. Light from a CW frequency-doubled Nd:YAG laser, at 532 nm (green), pumps a KTP (Potassium Titanyl Phosphate) crystal inside a Fabry-Perot cavity. Prior to pumping the OPO, this beam is filtered in a mode-cleaning cavity (bandwidth 2.4 MHz), to ensure it is in a coherent state for frequencies above 15 MHz. The OPO cavity is asymmetric: green light has only one input-output coupler, and the generated infrared light couples into and out of the cavity through the other mirror. The first mirror has moderate reflectivity at 532 nm ($R$=69.4%) and high reflectivity at 1,064 nm ($R > 99.8$%); the second has high reflectivity at 532 nm ($R > 99.8$%) and transmits a small fraction of light at 1,064 nm ($R$=96.0%). Above the oscillation threshold, bright, narrowband (~ 100 kHz linewidth), and frequency tunable twin beams are emitted by the OPO and their quadrature fluctuations are measured by reflecting them on quasi-resonant empty analysis cavities (*21*) prior to detection on high quantum efficiency photodiodes (overall detection efficiencies are 65 % for the pump beam and 87 % for the twins, accounting for losses in the beam paths). Any quadrature of any given field can be measured by scanning the resonance frequency of the respective analysis cavity. In this way, the full three-field covariance matrix of the system, with terms of the form $<\delta X_i(\Omega)\delta X_j(-\Omega)>$ and $<\delta X_i(\Omega)\delta Y_j(-\Omega)>$ (where $X$ and $Y$ represent the field amplitude or phase quadratures, $i,j$=0,1,2 and $\Omega$ is the analysis frequency), is directly measured, as previously reported (*22*).

A crucial improvement in the setup was needed in order to obtain tripartite entanglement. Thermal phonons present in the non-linear crystal generate excess phase noise and hinder the measurements, as recently described (*23*). Thus, we needed to cool down the crystal to temperatures below freezing. In order to avoid moisture on the crystal surfaces, a vacuum chamber was built around the OPO. By thermoelectric cooling, we reach -23$^\circ$C, temperature at which most of our data were obtained.

For Gaussian states, the complete information about the state is available from the covariance matrix. By inspecting higher-order moments (up to 10) from the measured photocurrents, we verified that within the experimental precision the three-field state from the above-threshold OPO is indeed Gaussian. We can thus verify a necessary and sufficient criterion for tripartite entanglement of Gaussian states: the positivity under partial transposition (PPT) (*24*). If one party is separable from the rest, the full density matrix remains positive under partial transposition with respect to that party. As shown by Simon (*24*), partial transposition for Gaussian states is equivalent to inverting the sign of one of the field's quadratures. The positivity can then be checked by evaluating the symplectic eigenvalues of the transposed matrix. The state is separable if and only if all symplectic eigenvalues are greater than or equal to one (*24*).

Our experimental demonstration of the full three-field inseparability is presented in Fig. 2. We characterize the OPO for several values of the pump power, relative to the threshold power ($\sigma = P/P_{th}$). For each value of $\sigma$, the covariance matrix is directly measured and the smallest symplectic eigenvalue under partial transposition for each field is evaluated. To assess its uncertainty, a Monte Carlo simulation is realized. Each term of the covariance matrix is randomly chosen within a Gaussian distribution having as standard deviation the experimental error (it also takes into account the error in the shot noise calibration, which is 0.6%). Ten thousand different covariance matrices compatible with the experimental values are tested for entanglement, producing well-behaved distributions (smooth and single-peaked) of symplectic eigenvalues thus providing the standard deviations of the respective distributions. Full inseparability is demonstrated for $\sigma$ ranging from 1.1 to 1.6, and for crystal temperatures below -10 $^o$C. In terms of the van Loock and Furusawa inequalities, which provide only a sufficient criterion, entanglement is also verified: $V_0 = 1.35 \pm 0.02$, $V_1 = 1.93 \pm 0.02$, and $V_2 = 1.93 \pm 0.02$, for $\sigma = 1.24$. This is the first direct generation of continuous-variable entanglement between more than two sub-systems and we take advantage of this approach by entangling fields of different wavelengths. Continuous-variable multipartite entanglement is classified in terms of full or partial inseparability (*25*). The phonon-induced phase noise in the OPO enables transitions from full inseparability to partial inseparability (the pump field becomes separable) just by tuning a single parameter, the crystal temperature.

Losses are always a concern in communications systems. It is important to know how robust the tripartite entanglement is. It is well known that all terms in the covariance matrix should linearly approach the SQL as losses increase (*26*). This is an important verification that the detection system is working properly. Although a squeezed state remains squeezed under finite linear losses, we find that tripartite entanglement behaves differently: we observe disentanglement of the pump field from the twins, for finite losses (Fig. 3). This new effect is reminiscent of so-called entanglement sudden death (*6,27*), observed in discrete-variable systems and predicted to occur in continuous-variable systems as well (*28,29*). For a pair of entangled qubits, each interacting with an independent environment, even though single-qubit coherence decays exponentially, entanglement can be lost at a finite time. Likewise, even though squeezing of individual beams tends only asymptotically to the SQL as a function of linear losses, we already observe disentanglement for finite losses. This can be harmful for applications to quantum communications, as in information networks. On the other hand, we observe that it is not present for all values of $\sigma$, as can be seen in Fig. 3, an important result for our system. Robustness of entanglement with respect to channel losses depends on $\sigma$, indicating that the entangled states are not of the same nature, albeit being always Gaussian. Further elaboration is under way.

The direct generation of higher-order continuous-variable entanglement demonstrated here already presents a novel fundamental feature, disentanglement for finite linear losses. This serves as a reminder that much is still unknown about the nature and dynamics of quantum entanglement, warranting further investigations. The relatively small violations of the entanglement criteria found here can be improved by better OPO cavity optics and further cooling of the non-linear crystal. Another possibility, currently under consideration, is to use different wavelengths. The OPO pump field can be at 780 nm, so as to interact resonantly with Rb atoms, generating output fields at wavelengths close to 1,560 nm. Quantum information stored in an atomic sample can thus be transferred to a wavelength suitable for propagation in low-loss optical fibers. Future quantum information networks may connect systems of different natures by multi-colored entangled light.

**Figure Legends**

**Fig. 1.** Experimental setup. The pump beam is provided by a frequency-doubled Nd:YAG laser. A filter cavity is used to ensure the field is in a coherent state for analysis frequencies above 15 MHz. An harmonic splitter (HS) allows the use of the fundamental frequency of the Nd:YAG laser for alignment purposes and Standard Quantum Limit (SQL) calibration. The pump reflected from the OPO cavity is directed to an analysis cavity, by a Faraday rotator (FR) and polarizing beam splitter cube (PBS). The KTP crystal can be thermoelectrically cooled down to -23 $^o$C, inside a vacuum chamber that prevents moisture on the crystal surfaces. The infrared twin beams are separated by another PBS and each is directed to an analysis cavity, which enables independent quadrature measurements on each beam. The fields reflected from each cavity are detected and the high-frequency component ($\Omega$ = 21 MHz, within a bandwidth of 600 kHz) of the photocurrent is treated in a demodulating chain.

**Fig. 2.** The symplectic eigenvalues, extracted from the measured covariance matrices, are plotted as functions of $\sigma = P/P_{th}$. In A), we present the measurements made at +23 $^o$C. Three eigenvalues are plotted, each corresponding to partial transposition with respect to one of the beams. We observe that at room temperature $\nu_0$ (transposition of the pump, green circles) is greater than one, while $\nu_1$ (transposition of signal, red squares) and $\nu_2$ (transposition of idler, blue triangles) are smaller than one for a large range of $\sigma$ values. Thus, at room temperature only bipartite entanglement between the twin beams exists. B) At -23 $^o$C we also observe $\nu_0$ drop below one, demonstrating the full inseparability of the three fields. The dashed lines are the usual theoretical prediction without taking into account the phonon noise. By taking into account the phonon noise (*22*), we obtain the solid lines, which agree well with the data. Error bars are standard deviations.

**Fig. 3.** We investigated the dependence of the symplectic eigenvalues as a function of linear losses imposed on the twin beams, by variable attenuators placed immediately prior to the photodetectors (pump beam losses had little effect). Color conventions are the same as in Fig. 2. In A) ($\sigma = 1.14$) and B) ($\sigma = 1.17$), we observe that the pump beam becomes separable from the twins for finite losses (transmittance near 0.6 and 0.4, respectively); on the right hand side, the situation is sketched: owing to the linear losses imposed on the twin beams, entanglement is lost; C) $\sigma = 1.40$ : there is no disentanglement for finite losses, all three fields remain inseparable until total loss, as sketched on the right hand side. Data were obtained at -10 °C. The solid curves are extracted from the covariance matrices obtained with no attenuation by calculating the effect of losses on each of their elements.

**Fig. 1**

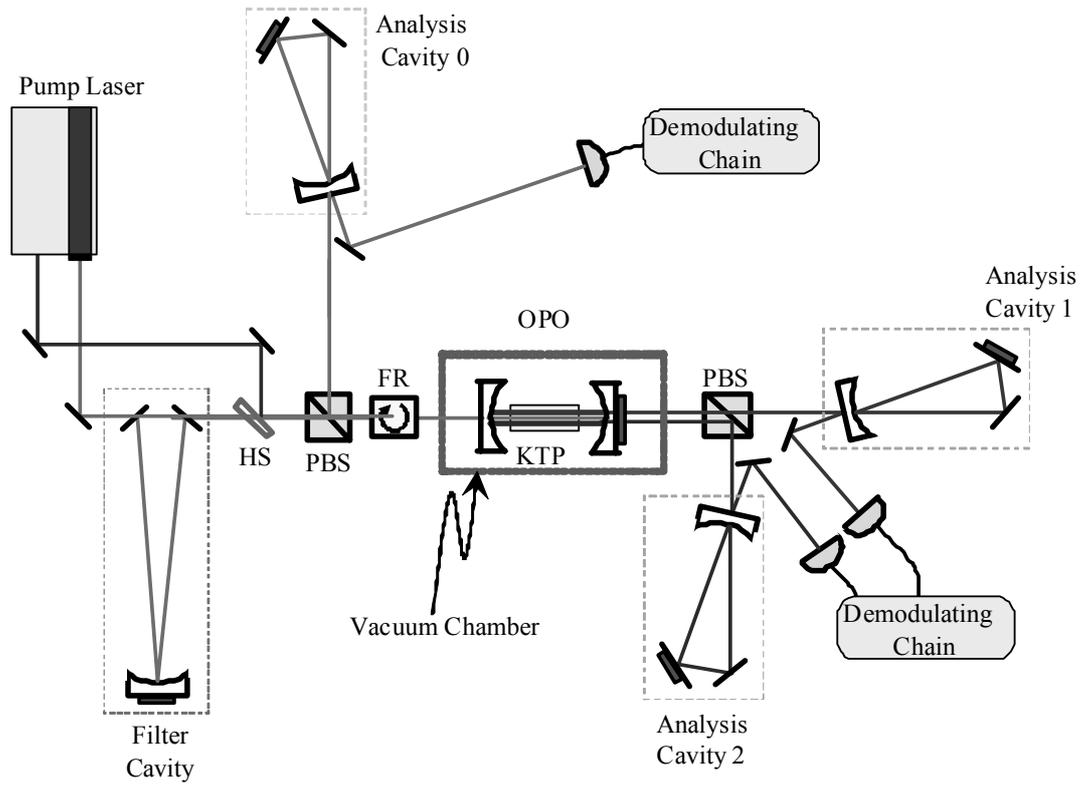

**Fig. 2.**

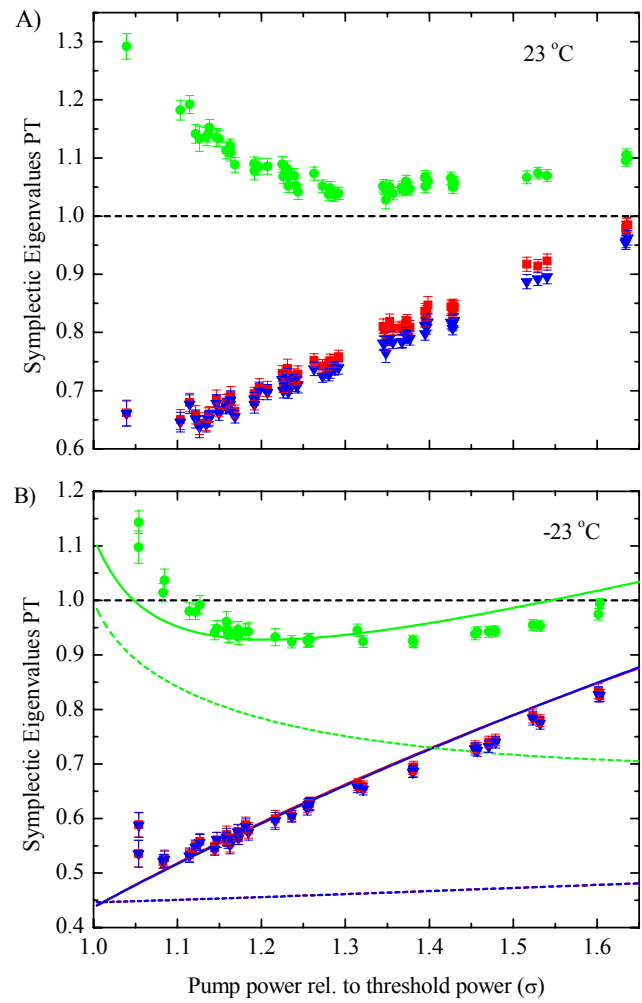

**Fig. 3.**

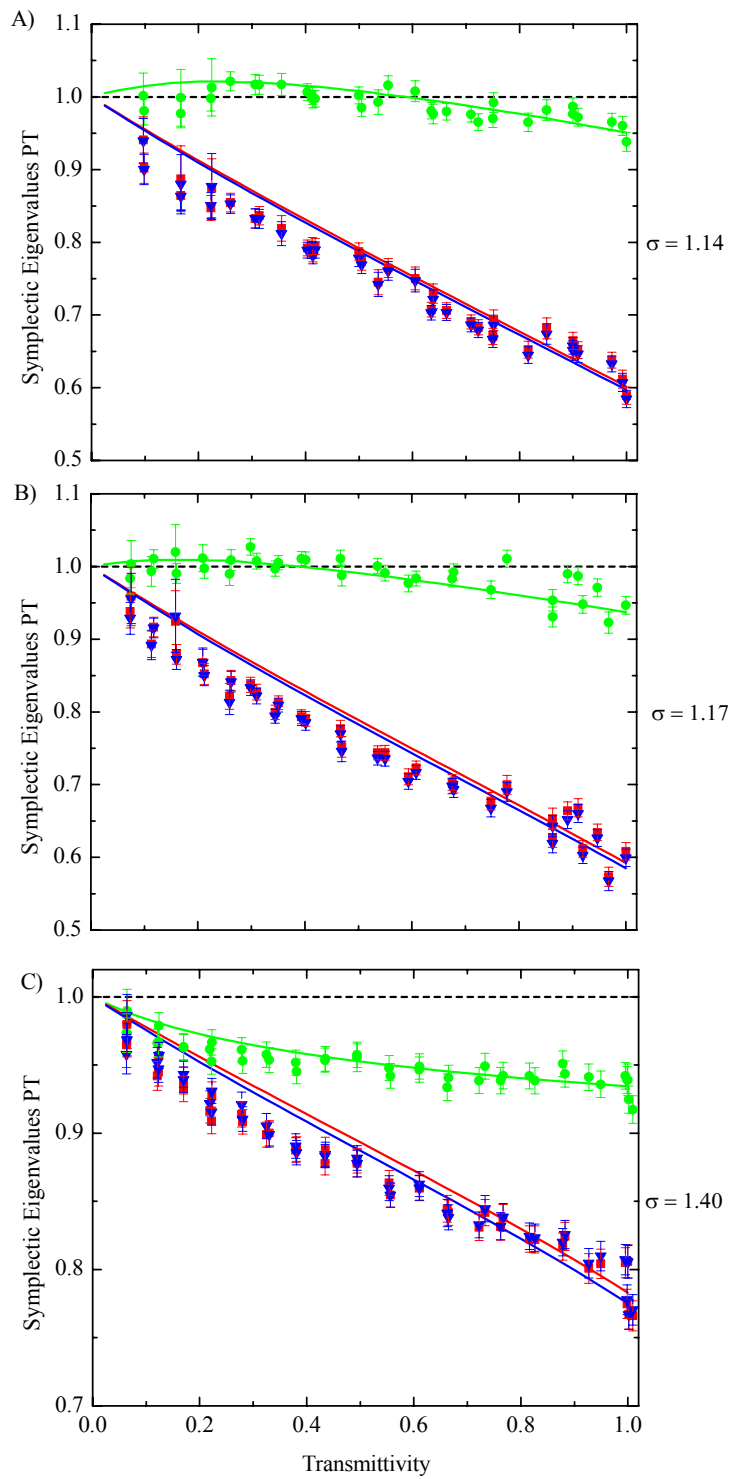
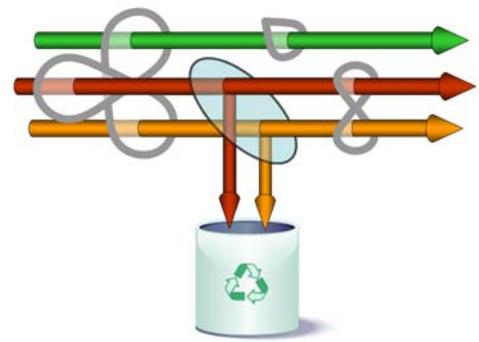
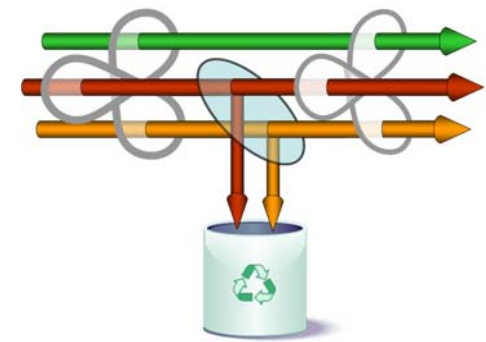